%% file: huber.tex
\documentclass{article}

\usepackage{spconf,amsmath,graphicx,cite,bm,amssymb,amsthm}

\usepackage[ruled,vlined]{algorithm2e}

\include{amiw_commands}

\begin{document}

\ninept

\title{Compressed matched filter for non-Gaussian noise}

\name{Jakob Vovnoboy and Ami Wiesel\thanks{This work was supported by the Israeli Smart Grid consortium, and also in part by ISF Grant 786/11. Special gratitude to Amir Globerson}}
\address{The Rachel and Selim Benin School of Computer Science and Engineering,\\ The Hebrew University of Jerusalem}
\maketitle

\begin{abstract}
 We consider estimation of a deterministic unknown parameter vector in a linear model with non-Gaussian noise. In the Gaussian case, dimensionality reduction via a linear matched filter provides a simple low dimensional sufficient statistic which can be easily communicated and/or stored for future inference.
Such a statistic is usually unknown  in the general non-Gaussian case. Instead, we propose a hybrid matched filter coupled with a randomized compressed sensing procedure, which together create a low dimensional statistic. We also derive a complementary algorithm for robust reconstruction given this statistic. Our recovery method is based on the fast iterative shrinkage and thresholding algorithm which is used for outlier rejection given the compressed data. We demonstrate the advantages of the proposed framework using synthetic simulations.
\end{abstract}

\section{Introduction}
One of the most fundamental concepts in parameter estimation is sufficient statistics. These are functions of the  observations that summarize all the information associated with the parameter of interest\cite{fisher1922mathematical}. Sufficient statistics minimize the required storage and communication resources. They are task independent and are useful when the data has to be compressed for a future specific use. Their computation usually involves simple and low complexity operations that are suitable for high rate processing. A sufficient statistic for estimation of a deterministic unknown parameter vector in a linear model with Gaussian noise is the well known matched filter. This simple linear operation is a core ingredient of most radar and communication systems.

 \begin{figure}\center{
 \includegraphics[width=0.5\textwidth]{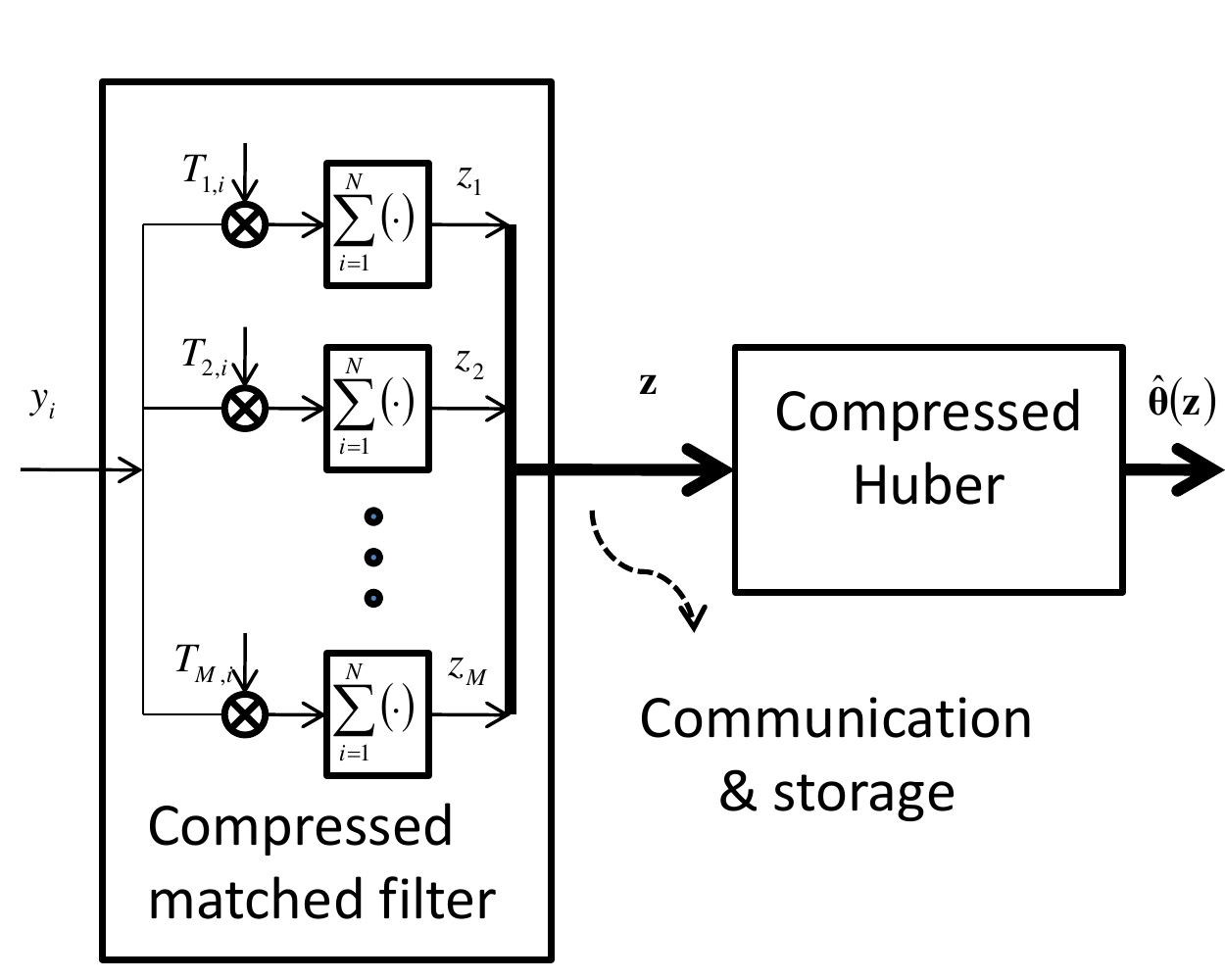}}
\caption{Block diagram of the CMF and CH.}
\end{figure}

  Many of the modern physical systems are better modeled as linear systems with non-Gaussian noise rather then Gaussian, this mainly due to impulsive noise phenomenons \cite{middleton1973man,wang1999robust,blum1999adaptive,blackard1993measurements,brockett1987nonlinear,kay1993fundamentals,huber1964robust,huber2011robust}. Typical noise characteristics include generalized Gaussian distributions, mixture distributions, impulsive models, and heavy tailed models.
In such scenarios, a low dimensional sufficient statistic is usually unknown, hence the classical matched filter is generally sub-optimal and more complicated non-linear operations are required \cite{chen1965matchedinnongauss}.

A common estimation technique in systems with non-Gaussian noise is to use non-linear element-wise limiters (also known as clippers) prior to the linear filter \cite{kay1993fundamentals}. However, this method is efficient only when the dynamic range of the data is small compared to the outliers.
  Another traditional solution in statistics is to resort to robust regression methods which works directly with the observed data, e.g., Huber's technique \cite{huber1964robust,huber2011robust}. This requires all the data to be stored for processing which can be quite expensive in terms of memory. In systems where the data have to be communicated for postprocessing the above is extremely limiting.
  Instead, the goal of this paper is to propose a compressed matched filter (CMF), namely a bank of approximate (and randomized) matched filters which compress the observations and contain most of the information in the data. Then, the output of the CMF can be easily stored or communicated for future use.  In addition, we derive a compressed Huber (CH) estimator which allows to reconstruct the unknown parameter vector using this compressed data.  A block diagram of CMF and CH is provided in Fig. 1.
  It is important to note that recovering the system parameters from the compressed data may still have the same complexity as in the uncompressed case. Therefore our main contribution is in lowering the amount of data transmitted from the receiver to the postprocessing unit.

Our framework is motivated by the theories of sparse recovery and compressed sensing (CS)\cite{baraniuk2007compressive, giannakis2011sparseoutliers}. Compressed sensing is a technique for reconstructing a high dimensional but sparse unknown vector from a small number of linear measurements. The basic approach is to use linear measurements with randomized coefficients and a recovery algorithm based on convex optimization with an $L_1$ norm. Sparse impulsive noise (or outliers) can be dealt with using the same framework \cite{laska2009sparslycorrupted, studer2011sparslycorrupted}.  A more advanced method known as LASSO extends the setting to noisy measurements \cite{tibshirani1996regression}. Recently, the problem was generalized to the estimation of a vector which is only partially sparse \cite{tirza:13}. All of these methods consider sparse signals and simple Gaussian noise. Interestingly, Fuchs (in \cite{fuchs1999inverse}) showed that Huber's robust regression can be expressed as the solution to a partially sparse model. The result is quite intuitive and can be interpreted as Gaussian noise contaminated by additional sparse and deterministic outliers. Fuchs had also suggested a generalization of the robust regression to the correlated noise case but did not present farther results. Our proposed systems is based on these ideas. CMF complements the classical matched filter with a few additional CS filters. CH estimates both the signal and the outliers by searching for a partially sparse solution that is consistent with the compressed data. Note that our frameworks is different than classical CS in two aspects: First, our sensing procedure is still mostly based on the matched filter. The additional randomized filter assist in detecting and eliminating the outliers. Second, unlike CS, our desired signal is dense. The sparsity is associated with the nuisance outliers.

The paper is organized as follows. We begin in Section \ref{sec_prob} by introducing the problem formulation. In Section \ref{sec_bounds} we consider the inherent performance limitations due to the compression. These bounds are computed assuming a clairvoyant estimator that can only be approximated in practice. In Section \ref{sec_matrix}, we address the choice of CMF using the theory of CS matrix design. In Section \ref{sec_alg}, we derive the CH algorithm which allows to reconstruct the unknown signal and as byproduct also part of the noise. This optimization is based on the joint Maximum a Posteriori and Maximum Likelihood (JMAP-ML) estimator \cite{yeredor2000joint} and ideas from Huber's regression \cite{huber1964robust}. The input to CH is low dimensional, but it processes high dimensional vectors in its internal computations. Thus, we also provide an efficient implementation of CH based on the fast iterative shrinkage and thresholding algorithm (FISTA) by \cite{beck2009fast}. Finally, in Section \ref{sec_sim} we illustrate the performance of our proposed methods using numerical simulations.

The following notation is used. The sets $\mathbb{R}^{n}$ and $\mathbb{R}^{n\times m}$ denote the set of length $n$ vectors and the set of size $n\times m$ matrices. The operator $\left\Vert \cdot\right\Vert _{p}$ denotes the $L_{p}$ norm. The superscript $\boldsymbol{X}^{T}$ and $\boldsymbol{X}^{-1}$ denotes the transpose and inverse operations. The subscript $x_{i}$ denote the $i$'s element in the vector $\boldsymbol{x}$. The Moore Penrose pseudoinverse of a matrix $\T$ is denoted by $\T^\dagger$.
 We denote the multivariate Gaussian distribution by $\mathcal{N}\left(\boldsymbol{\mu},\:\boldsymbol{\Sigma}\right)$ where $\boldsymbol{\mu}$ and $\boldsymbol{\Sigma}$ are the mean vector and the covariance matrix.

\section{Problem formulation}\label{sec_prob}
Consider a linear model
\begin{eqnarray}\label{model}
 \y=\H\thet+\n
\end{eqnarray}
where $\H\in\mathbb{R}^{N\times K}$ is a known matrix with $N\gg K$, $\thet\in\mathbb{R}^{K}$ is an unknown deterministic vector and $\n\in\mathbb{R}^{N}$ is a random vector with independent elements.
In the Gaussian case i.e.
\begin{eqnarray}
 n_i\sim {\mathcal{N}}\(0,\sigma_1^2\)\quad i=1,\cdots,N
\end{eqnarray}
It is well known that all the information in $\y$ about $\thet$ can be compressed by a linear matched filter
\begin{eqnarray}\label{Tcompress}
 \z=\T\y
\end{eqnarray}
  where $\T=\H^T$. i.e. $\z$ is a sufficient statistic of $\y$. Remarkably, the dimension of $\z$ is $K$ which is much smaller than $N$, and hence the compression. This is even true in the extreme continuous case in which $N$ is infinite but the dimension of $\z$ is still $K$ and depends only on the number of unknowns. Using $\z$ we can infer whatever we need about $\thet$ without storing $\y$. 

Our goal is to obtain a similar linear compression in the non-Gaussian case. Specifically, we assume
that the marginal distribution of each element in $\n$ is an {\it $\epsilon$-contaminated Gaussian model}
 \begin{eqnarray}\label{mixture}
 n_i\sim \(1-\epsilon\){\mathcal{N}}\(0,\sigma_1^2\)+\epsilon{\mathcal{G}}\quad i=1,\cdots,N
\end{eqnarray}
where $\epsilon>0$ is a known small contamination ratio parameter, and $\mathcal{G}$ is some symmetric distribution, typically unknown and referred to as outlier distribution. In this case, a low dimensional sufficient statistic is usually  unknown. Thus, we seek an approximate compression procedure. We will design a compression matrix $\T$ as in (\ref{Tcompress}) of size $M\times N$ where $K\leq M\ll N$, that summarizes as much information on $\thet$ as possible. Then, given the compressed $\z$ we will derive a computationally efficient algorithm for estimating the unknown parameter $\thet$.


\section{Performance bounds}\label{sec_bounds}
It is instructive to begin with two simplified problems which provide inherent performance bounds and explains our methodology. For this purpose, we assume a Gaussian mixture noise distribution meaning that the outliers are distributed normally, i.e. $\mathcal{G}=\mathcal{N}\(0,\sigma_2^2\)$ where $\sigma_{2}^{2}$ $\(\gg\sigma_{1}^{2}\)$ is a known constant, and consider oracle estimators which somehow know the locations of the outliers in $\n$. First, we consider the uncompressed case in which $\T=\I$. Under this assumption, the conditional distribution of the observations is
\begin{eqnarray}
 \z\sim {\mathcal{N}}\(\H\thet,\D\)
\end{eqnarray}
where $\D$ is a diagonal matrix with the variances of $\n$. Roughly, $\(1-\epsilon\) N$ of its diagonal elements are equal to $\sigma_1^2$ and the other $\epsilon N$ elements are equal to $\sigma_2^2$. This is a simple Gaussian linear model and the optimal estimator is  a Weighted Least Squares (WLS) \cite{kay1993fundamentals}
\begin{eqnarray}
 \hat\thet_{\text{no compression}}&=&\arg\min_{\thet}\;\|\y-\H\thet\|_{\D^{-1}}^2\\
 &=& \(\H^T\D^{-1}\H\)^{-1}\H^T\D^{-1}\z
\end{eqnarray}
where we use a weighted norm defined as
\begin{eqnarray}
 \|\x\|^2_{\W}=\x^T\W\x.
\end{eqnarray}
Its mean squared error is then given by
\begin{eqnarray}\label{unc_mse}
 {\text{MSE}}_{\text{no compression}}=\EE{\(\H^T\D^{-1}\H\)^{-1}}
\end{eqnarray}
where the expectation is with respect to the randomness in $\D$. This is not a general performance bound, as we have assumed a specific noise distribution but it is quite close if $\sigma_2^2$ is indeed the variance of the outliers. Any compression will probably increase the error, and our goal is to get as close as possible to this error with the smallest possible value of $M$.

In the compressed case (again, with known locations of the outliers and Gaussian outliers), the distribution of the observations is
\begin{eqnarray}
 \z\sim {\mathcal{N}}\(\T\H\thet,\T\D\T^T\)
\end{eqnarray}
where we condition on both $\T$ and $\D$ which are statistically independent. This too is a simple Gaussian linear model solved via a WLS
\begin{eqnarray}\label{oracle}
 \hat\thet_{\text{oracle}}=\(\H^T\T^T\!\(\T\D\T^T\)^{-1}\!\!\!\T\H\)^{-1}\!\!\!\!\!\H^T\T^T\(\T\D\T^T\)^{-1}\!\!\!\z
\end{eqnarray}
Its mean squared error is given by
\begin{eqnarray}\label{oracle_mse}
 {\text{MSE}}_{\text{oracle}}=\EE{\(\H^T\T^T\(\T\D\T^T\)^{-1}\T\H\)^{-1}}
\end{eqnarray}
where the expectation is with respect to the randomness in both $\H$ and $\D$. In practice, it is  impossible to implement the above oracle. However, it suggests a natural two step approach: first, detect the location of the outliers, then use an approximate oracle assuming these locations are exact. Furthermore, these MSEs are reasonable performance bounds that any practical estimator should be compared to.

\section{Compressed matched filter for non-Gaussian noise}\label{sec_matrix}
The first part of our design is the choice of the sensing matrix $\T$ which defines CMF. Unfortunately, it is not completely clear what is the optimal criterion for the design and/or how to numerically optimize it.
On the one hand, from the signal perspective, we would like $\T$ to be close to the matched filter. At the least we need to ensure that its columns span the columns of $\H$.
On the other hand, we need $\T$ to give some response to the noise shape.
  Here we can use some insights from the compressed sensing field by looking at a sparse model which is close to ours. Specifically, we can look at a linear system with Gaussian noise and deterministic outliers. Hence, setting $\n=\bnu+\u$ where $\bnu$ is a random vector with independent ${\mathcal{N}}\(0,\sigma_1^2\)$ variables, and $\u$ is a deterministic sparse vector into (\ref{model} results in the following model
  \begin{eqnarray}\label{outlier_model}
  \y=\H\thet+\bnu+\u
 \end{eqnarray}
 CS theory hints that we can use a random matrix to encode the sparsity of $\u$.
  Therefore, to address both criteria we propose the following simple structure:
\begin{eqnarray}
 \T=\[
\begin{array}
 {c} \H^T \\ \W\P
\end{array}
\]
\end{eqnarray}
where $\W\in\mathbb{R}^{M-K\times N}$ is a matrix with independent and identically distributed (i.i.d.)  ${\mathcal{N}}\(0,1\)$ elements and
\begin{eqnarray}
 \P=\I-\H\(\H^T\H\)^{-1}\H^T
\end{eqnarray}
is a projection matrix onto the null space of $\H$. This choice guarantees that we will always be better or equal to the naive Gaussian matched filter which is exactly the first $K$ rows. The rest of the rows randomly span as much as possible from the remaining space.


 \section{Compressed Huber}\label{sec_alg}

 The second part of our framework is the CH algorithm which estimates $\thet$ given $\z$ for a fixed $\T$. Ideally, we would like to find the parameter $\thet$ that maximizes the likelihood of $\z$. But this vector is a high dimensional mixture of many non-Gaussian random variables, and its distribution is hard to analyze. Instead, we propose to estimate both $\thet$ and $\n$ simultaneously. Statistically speaking, we jointly seek for $\thet$ via a maximum likelihood approach and for $\n$ via a maximum a posteriori approach (see \cite{yeredor2000joint} for more details on JMAP-ML estimation):
\begin{eqnarray}\label{jmlmap}
\begin{array}
 {ll} \min_{\thet,\n} & \sum_{i=1}^N \phi\(n_i\) \\
 {\text{s.t.}} & \z=\T\(\H\thet+\n\)
\end{array}
\end{eqnarray}
 where $\phi\(\cdot\)$ is the negative-log-posterior distribution of $n_i$ as described in (\ref{mixture}). Because the distributions of $\mathcal{G}$ and as a result the distribution of $\n$ are generally unknown, we can not calculate $\phi\(\cdot\)$ directly. Hence, we have to use some robust objective function which will be indifferent to the specific distribution of the outliers. Such is the Huber's loss function which was proven to be optimal in the uncompressed case (in the minimax sense) \cite{huber2011robust}.
 Together, our reconstruction algorithm is the solution to
\begin{eqnarray}\label{jmlmaphuber}
\begin{array}
 {ll} \min_{\thet,\n} & \sum_{i=1}^N \rho_h\(n_i\) \\
 {\text{s.t.}} & \z=\T\(\H\thet+\n\)
\end{array}
\end{eqnarray}
  where
  \begin{eqnarray}
 \rho_h(n)=
\left\{\begin{array}
 {ll} n^2 & |n|<h \\ 2h|n|-h^2 & |n|>h
\end{array}\right.
\end{eqnarray}
and $h$ is calculated from
\begin{eqnarray}
 \frac{\sigma}{h}\psi\(\frac{h}{\sigma}\)-Q\(\frac{h}{\sigma}\)=\frac{\epsilon}{2\(1-\epsilon\)}
\end{eqnarray}
with $\psi(x)=\frac{1}{\sqrt{2\pi}}e^{-\frac{x^2}{2}}$ and $Q(t)=\frac{1}{\sqrt{2\pi}}\int_t^\infty e^{-\frac{x^2}{2}}dx$.
 The above is a convex minimization that can be efficiently solved using off-the-shelf optimization packages, e.g., CVX \cite{grant2008cvx}.

Remarkably, \cite{fuchs1999inverse} showed that Huber's function can be expressed as:
\begin{eqnarray}
 \rho_h(n)=\min_u\; (u-n)^2+2h|u|
\end{eqnarray}
Plugging this expression into (\ref{jmlmaphuber}) yields
\begin{eqnarray}\label{jmlmaphuber_fuchs}
\begin{array}
 {ll} \min_{\thet,\n,\u} & \|\n-\u\|^2+2h\|\u\|_1 \\
 {\text{s.t.}} & \z=\T\(\H\thet+\n\)
\end{array}
\end{eqnarray}
Then by solving explicitly for $\n$ we can derive the following equivalent problem
\begin{eqnarray}\label{lasso_comp}
 \min_{\thet,\u} \left\|\z-\T\[\H\thet+\u\]\right\|^2_{\(\T\T^T\)^{-1}}+2h\|\u\|_1
\end{eqnarray}
This formulation provides an interesting observation. The solution of  (\ref{lasso_comp}) can be interpreted as an estimator to the deterministic sparse outlier model in (\ref{outlier_model}). It can be seen that the first term in (\ref{lasso_comp}) is a standard WLS objective whereas the second term penalizes vectors $\u$ which are not sparse.


The above formulation is also useful from a numerical perspective. Note that $\z$ is compressed and low dimensional, but the internal variable $\u$ is of length $N$ and therefore the optimization requires a large scale numerical algorithm. For this purpose, we utilize the well known FISTA solver due to \cite{beck2009fast}.
 First, we notice that $\hat\thet$ can be solved explicitly
\begin{eqnarray}\label{thta_exp}
\hat\thet=\Q\(\z-\T\u\)
\end{eqnarray}
where $\Q=\(\H^T\T^\dagger\T\H\)^{-1}\H^T\T^\dagger$. Then by substituting it to (\ref{lasso_comp}) we get a classical LASSO problem
\begin{eqnarray}\label{lasso_comp_only_u}
 \min_{\u} \left\|\(\I-\T\H\Q\)\(\z-\T\u\)]\right\|^2_{\(\T\T^T\)^{-1}}+2h\|\u\|_1
\end{eqnarray}
which can be solved efficiently by FISTA.
For convenience of notation, we also define the shrinkage and thresholding operator
\begin{eqnarray}
 {\mathcal{T}}_\alpha(x)=\max\{|x|-\alpha,0\}\sign{x}
\end{eqnarray}
Summing the above, a pseudocode for solving CH in (\ref{lasso_comp}) using FISTA is provided in Algorithm 1.

 \begin{algorithm}
\KwIn{$\T$, $\H$, $\z$}
\KwOut{$\hat\thet$,$\hat\u$}
$\S=\(\I-\H\Q\)^T\(\T\T^T\)^{-1}\(\I-\H\Q\)$\\
$L=2\lambda_{\max}\(\T^T\S\T\)$\\
Step 0: Take $t_1=1$ and \\
$\qquad \u_1=\tilde\u=\0$\\
Step $k$: ($k\geq 0$) Compute\\
$\qquad \e_k=\z-\T\tilde\u_k$\\
$\qquad \u_k={\mathcal{T}}_{\frac{2h}{L}}\(\tilde\u_k-\frac{2}{L}\T^T\S\e_k\)$\\
$\qquad t_{k+1}=\frac{1+\sqrt{1+4t_k^2}}{2}$\\
$\qquad\tilde\u_{k+1}=\u_k+\frac{t_k-1}{t_{k+1}}\(\u_k-\u_{k-1}\)$\\

Return: $\hat\u=\u_k$ and $\hat\thet=\Q\(\z-\T\u_k\)$
\caption{FISTA implementation of CH}
\label{bisection}
\end{algorithm}

After computing CH, we propose to fine tune the estimate. By examining the optimal $\u$ we (approximately) detect the locations of the outliers
\begin{eqnarray}
\hat{\mathcal{I}}_{out}=\left\{ i|\ |\hat{u}_{i}|>\sigma_{1}\right\}
\end{eqnarray}
Then we estimate their variance
\begin{eqnarray}
\hat{\sigma}_2^2=\frac{1}{\left|\hat{\mathcal{I}}_{out}\right|}\underset{i\in\hat{\mathcal{I}}_{out}}{\sum \hat{u}_{i}^{2}}
\end{eqnarray}
  recover the diagonal covariance matrix of $\n$ denoted by $\hat\D$ and finally compute $\hat\thet_{\text{oracle}}$ in (\ref{oracle}) replacing the true $\D$ with its estimate $\hat\D$. We denote this second phase as AWLS for approximate WLS.

 \section{Numerical results}\label{sec_sim}
 To demonstrate the advantages of CMF and CH we present simulation results in a simple signal processing application. We consider the estimation of amplitudes and phases of $K/2$ sinusoids with known frequencies contaminated by several non-Gaussian noises.
  Specifically, we define $K=10$, $N=500$, $\epsilon=1\%$, $\sigma_1^2=1$ and $\sigma_2^2=500$. We express the sinusoids in linear form by defining  $ \H=\[\;\H^c\;\H^s\;\]$ where $\H^c_{i,n}=\cos\(2\pi f_i\(n-1\)\)$, $\H^s_{i,n}=\sin\(2\pi f_i\(n-1\)\)$, $i=1,\cdots, K/2$ and $n=1,\cdots,N$. The frequencies are $f_1=.1$, $f_2=.2$, $f_3=.3$, $f_4=.35$, and $f_5=.4$. The true amplitudes are all unit and the data is contaminated with a Gaussian mixture noise i.e. $\mathcal{G}_i\sim{\mathcal{N}}\(0,\sigma_2^2\)$. The data is compressed using CMF and the system parameters are estimated using the suggested algorithms.
   
  Fig. 2 presents the estimation mean squared errors averaged over the realizations of the noise $\n$ as a function of compression ratio $M/N$. For comparison, the errors are bounded below by computing (\ref{unc_mse}) (NO COMPRESSION) and above by computing (\ref{oracle_mse}) with $\T=\H^T$ (FULL COMPRESSION). In between is the ORACLE using the proposed CMF with randomized versions of $\T$. Our proposed estimators are denoted by CH and AWLS.
   It is easy to see the advantages of CH which closes $90\%$ the performance gap with only a quarter of the complete measurements (i.e. four fold compression). AWLS is even better and achieves the same performance with a higher compression. On the downside, the simulation suggest that there may still be room for improvement. Neither CH nor AWLS succeed in achieving the ORACLE performance that knows the locations of the outliers. Similar results were obtained for Laplace distributed outliers with the same variance.

Fig. 3 presents the estimation (using CH) mean squared errors averaged over the realizations of the noise $\n$ as a function of $N$ for several compression ratios. As can be seen the estimation error is monotonic and asymptotically tends to zero inversely proportional to $N$ $\(\mathrm{MSE}\propto N^{-1}\)$. Thus, suggesting asymptotic consistency of the estimator. It is also worth noting that asymptotically the ratio between estimation MSE's for different compression is constant.
    \begin{figure}\center{
   \includegraphics[width=0.5\textwidth]{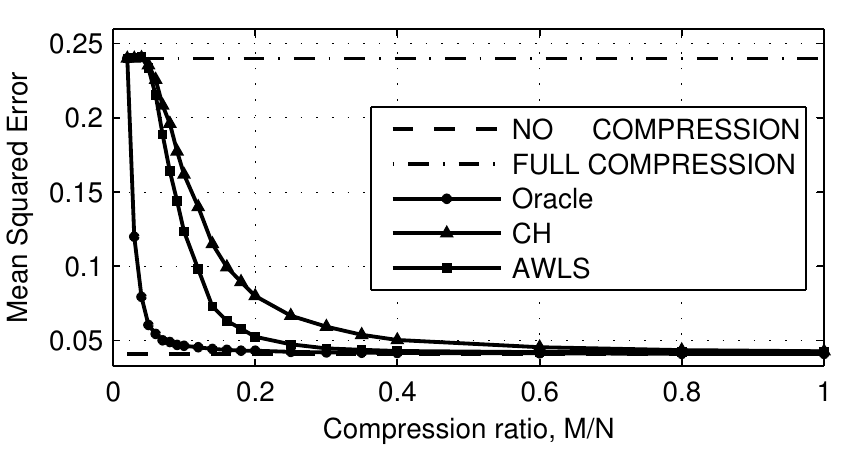}}
   \caption{Estimation mean squared error as a function of compression ratio $M/N$.}
   \end{figure}

    \begin{figure}\center{
 \includegraphics[width=0.5\textwidth]{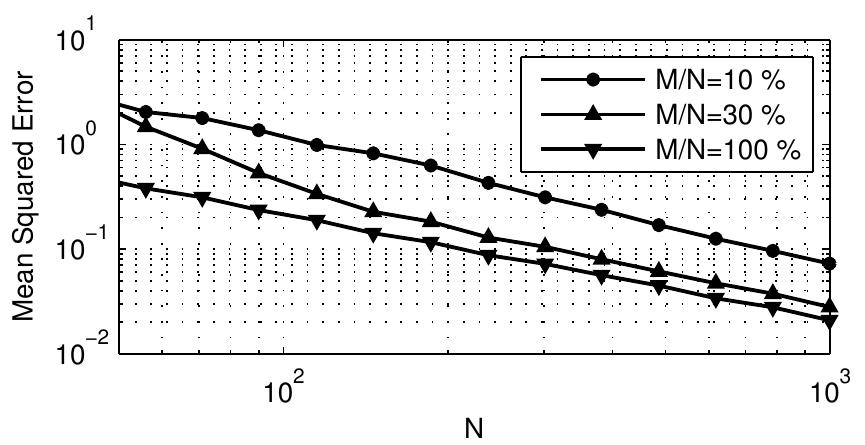}}
\caption{Estimation (using CH) mean squared error as a function of $N=50,\cdots,1000$. }
\end{figure}

\section{Conclusions and future work}
In this paper we have presented a simple compression scheme for a linear system without major information loss. We have also developed a fast recovery scheme for the compressed data.
Combination of the two methods were shown, by simulations, to recover the system parameters using approximately four fold compression with no significant loss in MSE. Additional research is needed to optimize the compression matrix and finding more efficient recovery algorithms  or providing a tighter lower bound for them.
\bibliographystyle{plain}
\bibliography{ssp}

\end{document}

%% file: amiw_commands.tex
\renewcommand{\(}{\left(}
\renewcommand{\)}{\right)}
\renewcommand{\[}{\left[}
\renewcommand{\]}{\right]}

\newcommand{\n}{\mathbf{n}}

\renewcommand{\S}{\mathbf{S}}

\newcommand{\thet}{\bm{\theta}}

\newcommand{\bnu}{\bm{\nu}}

\newcommand{\y}{\mathbf{y}}
\renewcommand{\H }{\mathbf{H}}
\newcommand{\z}{\mathbf{z}}

\newcommand{\W}{\mathbf{W}}

\newcommand{\0}{\mathbf{0}}
\newcommand{\D}{\mathbf{D}}

\newcommand{\x}{\mathbf{x}}
\newcommand{\I}{\mathbf{I}}

\renewcommand{\P}{\mathbf{P}}

\newcommand{\T}{\mathbf{T}}

\renewcommand{\u}{\mathbf{u}}

\newcommand{\Q}{\mathbf{Q}}

\newcommand{\e}{\mathbf{e}}

\newcommand{\EE}[1]{{\rm{E}}\left\{#1\right\}}

\renewcommand{\arg}[1]{{\rm{arg}}#1}

\newcommand{\sign}[1]{{\rm{sign}}\left\{#1\right\}}